\documentclass[a,preprint,pdf]{iucr}
\pdfoutput=1

\usepackage[utf8]{luainputenc}
\usepackage[version=4]{mhchem}

\usepackage{bm,xspace,amsmath,amssymb,siunitx,graphicx}

\DeclareSIUnit\pixel{px}

\newcommand{\rperp}{\ensuremath{\vec{r}_\perp}}
\newcommand{\kperp}{\ensuremath{\vec{k}_\perp}}
\newcommand{\ee}{\ensuremath{\mathrm{e}}}
\newcommand{\ii}{\ensuremath{\mathrm{i}}}
\newcommand{\dd}{\ensuremath{\mathrm{d}}}
\newcommand{\laplace}{\op{\bigtriangleup}}
\newcommand{\Lz}{\ensuremath{\op{L}_z}\xspace}
\newcommand{\expect}[1]{\ensuremath{\left\langle #1 \right\rangle}}

\newcommand{\op}[1]{\ensuremath{\hat{#1}}}
\renewcommand{\vec}[1]{\ensuremath{\boldsymbol{#1}}}
\newcommand{\mat}[1]{\ensuremath{\boldsymbol{#1}}}
\DeclareMathOperator{\FT}{\mathcal{F}}

\newcommand{\SiN}{\ce{Si3N4}\xspace}
\newcommand{\FeB}{\ce{Fe_{0.8}B_{0.2}}\xspace}
\newcommand{\Pt}{\ce{Pt}\xspace}
\newcommand{\Ang}{{\AA}ngström\xspace}

\begin{document}

\title{Elastic propagation of fast electron vortices through amorphous materials}
\shorttitle{Electron vortices in amorphous materials}

\cauthor[a]{Stefan}{Löffler}{stefan.loeffler@tuwien.ac.at}{}
\author[b]{Stefan}{Sack}
\author[a,b]{Thomas}{Schachinger}
\aff[a]{University Service Centre for Transmission Electron Microscopy, TU Wien, Wiedner Hauptstraße 8-10/E057-02, \city{Wien}, \country{Austria}}
\aff[b]{Institute of Solid State Physics, TU Wien, Wiedner Hauptstraße 8-10/E138-03, \city{Wien}, \country{Austria}}


\maketitle

\begin{synopsis}
This article describes the scattering behavior of electron vortices inside amorphous samples. It focuses on the vortex purity, net angular momentum transfer, and statistical variations due to random beam and atom positions.
\end{synopsis}

\begin{abstract}
In this work, we study the elastic scattering behavior of electron vortices when propagating through amorphous samples. We use a formulation of the multislice approach in cylindrical coordinates to theoretically investigate the redistribution of intensity between different angular momentum components due to scattering. To corroborate and elaborate on our theoretical results, we perform extensive numerical simulations on three model systems (\SiN, \FeB, \Pt) for a wide variety of experimental parameters to quantify the purity of the vortices, the net angular momentum transfer, and the variability of the results with respect to the random relative position between the electron beam and the scattering atoms. These results will help scientists to further improve the creation of electron vortices and enhance applications involving them.
\end{abstract}

\section{Introduction}

The study of electron vortex beams (EVBs) is a highly active field of research in the context of transmission electron microscopy (TEM). The fact that these EVBs carry orbital angular momentum (OAM) has lead to many demonstrated and proposed applications ranging from the measurement of magnetic properties with atomic resolution \cite{N_v467_i7313_p301,PRB_v89_i_p134428,U_v136_i_p81,ASaCI_v2_p5,U_v179_i_p15} over the study of the dynamics of Landau states \cite{NC_v5_i_p4586,U_v158_i_p17}, sample chirality \cite{PRB_v91_i9_p94112}, and symmetry properties of plasmon resonances \cite{Nc_v8_i_p14999} to the manipulation of nanoparticles \cite{AM_v25_i8_p1114}. Despite their huge potential and although their creation and propagation through vacuum is well-understood \cite{U_v111_i9-10_p1461,U_v115_i0_p21,U_v158_i_p17}, the knowledge of their propagation through matter is still somewhat lacking. This is especially surprising since earlier studies showed that elastic scattering in crystals can drastically change the orbital angular momentum of the beam \cite{ACA_v68_i_p443,MaM_v18_i_p711,PRA_v87_i_p33834}.

Particularly important --- and little investigated --- is the propagation of EVBs through amorphous materials. Firstly, such materials are used increasingly often for producing EVBs by means of specially designed phase masks \cite{NJoP_v16_i_p93039,U_v144_i_p26,APL_v104_i4_p43109}. Secondly, they are a common support, e.g. for nanoparticles. Thirdly, EVBs would allow techniques such as energy-loss magnetic chiral dichroism (EMCD) for measuring magnetic properties down to the nanoscale in crystalline samples to be applied also to amorphous materials. However, it is usually assumed that an as-produced, ideal vortex beam stays that way and propagates practically unperturbed through the sample. Whether or not that is the case and, if so, to what extent is studied in this work.

This paper is structured as follows: first, we give a brief overview over the theory in sec.~\ref{sec:theory}. To that end, we rewrite the multislice approach used throughout this work in a cylindrical coordinate system suitable for the analysis of EVBs. From that, we deduce some general statements about the propagation behavior of EVBs. In sec.~\ref{sec:sim}, we give a detailed account of the numerical simulations performed in this work. In sec.~\ref{sec:results}, the results of the numerical simulations are presented, which are subsequently discussed in sec.~\ref{sec:discussion}.

\section{Theory}
\label{sec:theory}

The starting point for describing the propagation of electrons through matter is Schrödinger's equation. Throughout this work, we will adopt a paraxial multislice approach \cite{Kirkland1998}. In this approach, the sample is cut into thin slices and the propagation of an electron wave function through slice $n$ is given by
\begin{equation}
	\psi_n(\rperp) = \ee^{\frac{\ii t_n}{2k_z} \laplace} \ee^{-\ii \sigma v_{z,n}(\rperp)} \psi_{n-1}(\rperp)
	\label{eq:prop_general}
\end{equation}
where $\rperp$ is the 2D coordinate vector in the $x$-$y$-plane perpendicular to the main propagation direction $z$, $\psi_{n-1}(\rperp)$ is the wave function incident on the $n$-th slice, $\psi_{n}(\rperp)$is the wave function exiting the $n$-th slice, $k_z$ is the $z$-component of the wave vector, $t_n$ is is thickness of the slice, $\laplace$ is the Laplace operator, $\sigma$ is the so-called interaction parameter, and $v_{z,n}(\rperp)$ is the electrostatic potential of the slice projected along the $z$ direction. In eq.~\ref{eq:prop_general}, the $\ee^{-\ii \sigma v_{z,n}(\rperp)}$ term describes (instantaneous) elastic scattering, while the $\ee^{\frac{\ii t_n}{2k_z}\laplace}$ term describes the free-space Fresnel propagation through the slice. To propagate through the entire sample, many such individual propagation steps have to be performed. Note that $\laplace$ and $v_{z,n}(\rperp)$ do not generally commute, so the exponentials cannot easily be reordered.

Here, we are primarily interested in the evolution of the different OAM components, so we expand the terms in eq.~\ref{eq:prop_general} into the eigenstates $\exp(\ii m \varphi)$ of the OAM operator $\Lz=-\ii\hbar\ \partial/\partial\varphi$:
\begin{align*}
	\psi_n(\rperp) &= \sum_m f_{n,m}(r) \ee^{\ii m \varphi} \\
	\ee^{-\ii\sigma v_{z,n}(\rperp)}  &= \sum_\mu V_{n,\mu}(r) \ee^{\ii \mu \varphi},
\end{align*}
with
\begin{align}
	f_{n,m}(r) &= \frac{1}{2\pi} \int_0^{2\pi} \psi_n(r, \varphi) \ee^{-\ii m \varphi} \dd\varphi \notag \\
	V_{n,\mu}(r) &= \frac{1}{2\pi} \int_0^{2\pi} \ee^{-\ii\sigma v_{z,n}(r, \varphi)} \ee^{-\ii \mu \varphi} \dd\varphi, \label{eq:Vmu}
\end{align}
where $(r, \varphi)$ denote the polar components of \rperp. In physical terms, $m$ denotes the topological charge of a vortex component with an OAM of $m\hbar$. With these definitions, eq.~\ref{eq:prop_general} reduces to
\begin{align}
	\sum_m f_{n,m}(r) \ee^{\ii m \varphi} &= \ee^{\frac{\ii t_n}{2k_z} \laplace} \sum_{m,\mu} V_{n,\mu}(r) f_{n-1,m}(r) \ee^{\ii (m+\mu) \varphi} \notag\\
	&= \ee^{\frac{\ii t_n}{2k_z} \laplace} \sum_{m} \left[ \sum_\mu V_{n,m-\mu}(r) f_{n-1,\mu}(r) \right] \ee^{\ii m \varphi} \notag\\
	&= \ee^{\frac{\ii t_n}{2k_z} \laplace} \sum_{m} g_{n,m}(r) \ee^{\ii m \varphi}
	, \label{eq:scattering_oam}
\end{align}
i.e., the elastic scattering transforms the set of radial components $\{f_{n-1,m}(r)\}_m \mapsto \{g_{n,m}(r)\}_m$.

The action of the Laplacian operator, i.e., the Fresnel-propagation between the slices, is best viewed in reciprocal space. There, the 2D Laplacian reduces to $k_\perp^2$ and the OAM distribution is maintained \cite{U_v115_i0_p21}, giving
\[
	\sum_m f_{n,m}(r) \ee^{\ii m \varphi} = \FT_{\kperp\to\rperp} \left[ \ee^{\frac{\ii t_n k^2}{2k_z}} \sum_{m} g_{n,m}(k) \ee^{\ii m \varphi_k} \right]
\]
with the Hankel transforms
\begin{align}
g_{n,m}(k) &= \ii^m \int_0^\infty g_{n,m}(r)J_m(k r)r\dd r \notag \\
f_{n,m}(r) &= \frac{1}{\ii^m} \int_0^\infty \ee^{\frac{\ii t_n k^2}{2k_z}} g_{n,m}(k)J_m(k r)k\dd k \label{eq:fn_after_prop}
\end{align}
where $(k, \varphi_k)$ are the polar coordinates of the vector \kperp, $\FT_{\kperp\to\rperp}$ denotes the 2D Fourier transform from reciprocal to real space, and $J_m$ is the Bessel function of first kind of order $m$. 

It can be seen that the redistribution of intensity between different OAM components happens due to the elastic scattering in the electrostatic potential $v_z$ (see eq.~\ref{eq:scattering_oam}), while during the Fresnel propagation, only the radial distributions evolve but no intensity is transferred between different OAM components. The potential scattering term can also be written in vector form as
\[
	\vec{g}_n(r) = \mat{V}_n(r) \cdot \vec{f}_{n-1}(r),
\]
where
\[
	(\mat{V}_n(r))_{m,m'} = V_{n,m-m'}(r)
\]
is a Toeplitz matrix.

There are several noteworthy points here.
First of all, scattering from a component $m$ to a component $m+\delta m$ takes place only if there exists some $r$ for which $V_{n,\delta m}(r)$ and $f_{n-1,m}(r)$ are both non-negligible. On the one hand, this reflects the obvious fact that only those areas of the potential affect the beam in which the beam intensity is non-vanishing. On the other hand, it also implies certain symmetry properties (see sec.~\ref{sec:symmetry}).

Secondly, one can expect $\delta m=0$ to be the dominant term for thin slices. This results from the fact that for thin slices, $v_z$ is small. Thus, the potential can be written in weak-phase-object approximation as
\[
	\ee^{-\ii\sigma v_{z,n}(\rperp)} \approx 1 - \ii\sigma v_{z,n}(\rperp),
\]
showing that there is a large constant term, which results in a large $\delta m = 0$ contribution.


\subsection{Symmetry Constraints}
\label{sec:symmetry}

Symmetry plays an important role in the scattering behavior of electron beams, especially in crystalline specimens. Even though in amorphous materials, the potential typically does not exhibit strict symmetries, it can still show certain ``approximate'' symmetries, i.e., atomic arrangements that deviate only slightly from a symmetric case. In fact, while in crystalline samples, symmetries typically only hold for certain special, high-symmetry points such as atomic columns and are severely broken if the electron beam is positioned off-column, the random distribution of atoms in amorphous systems means that the same symmetry properties hold in an approximate sense fairly independently of the beam position. Thus, a closer investigation of the symmetry constraints for OAM transfer seems worthwhile.

Here, we consider the inherently two-dimensional case in the plane perpendicular to the beam axis (i.e., in a slice). More precisely, we study the transformation properties of the potential scattering term $\exp(-\ii\sigma v_{z,n}(r, \varphi))$ under the point group $O(2)$, which contains rotations and reflections (as well as arbitrary combinations of them).

For the case of rotations, we assume that the potential has a $\nu$-fold rotational symmetry, i.e. $v_{z,n}(r, \varphi + 2\pi/\nu) = v_{z,n}(r, \varphi)$. Inserting this into eq.~\ref{eq:Vmu} yields
\begin{align*}
	& \int_0^{2\pi} \ee^{-\ii\sigma v_{z,n}(r, \varphi)} \ee^{-\ii \mu \varphi} \dd\varphi \\
	={}& \sum_{j=0}^{\nu-1} \int_{\frac{2\pi j}{\nu}}^{\frac{2\pi (j+1)}{\nu}} \ee^{-\ii\sigma v_{z,n}(r, \varphi)} \ee^{-\ii \mu \varphi} \dd\varphi \\
	={}& \int_{0}^{\frac{2\pi}{\nu}} \ee^{-\ii\sigma v_{z,n}(r, \varphi)} \ee^{-\ii \mu \varphi} \dd\varphi \cdot \sum_{j=0}^{\nu-1} \ee^{-2\pi\ii \frac{\mu}{\nu} j} \\
	={}& \begin{cases}
		0 & \mu/\nu \notin \mathbb{Z} \\
		\nu\int_{0}^{\frac{2\pi}{\nu}} \ee^{-\ii\sigma v_{z,n}(r, \varphi)} \ee^{-\ii \mu \varphi} \dd\varphi & \mu/\nu \in \mathbb{Z} \\
	\end{cases}
\end{align*}
using the summation formula for finite geometric series. Therefore, in the case of a $\nu$-fold rotational symmetry of the potential around the beam axis, $V_{n,\mu} \equiv 0\ \forall \mu \notin \nu\mathbb{Z}$, i.e., intensity can only be redistributed between OAM components which differ by an integer multiple of $\nu\hbar$.

For the case of reflections, we assume that the potential is symmetric with respect to a mirror line inclined by an angle $\varphi_0$ with respect to the $x$-axis, i.e. $v_{z,n}(r, \varphi_0 - \varphi) = v_{z,n}(r, \varphi_0 + \varphi)$. Inserting this into eq.~\ref{eq:Vmu} yields
\begin{align*}
	& \int_{\varphi_0-\pi}^{\varphi_0+\pi} \ee^{-\ii\sigma v_{z,n}(r, \varphi)} \ee^{-\ii \mu \varphi} \dd\varphi \\
	={}& \int_{0}^{\pi} \ee^{-\ii\sigma v_{z,n}(r, \varphi_0-\varphi)} \ee^{-\ii \mu (\varphi_0-\varphi)}  \dd\varphi \\
	& {} + \int_{0}^{\pi} \ee^{-\ii\sigma v_{z,n}(r, \varphi_0+\varphi)} \ee^{-\ii \mu (\varphi_0+\varphi)} \dd\varphi \\
	={} & 2 \ee^{-\ii \mu \varphi_0} \int_{0}^{\pi} \ee^{-\ii\sigma v_{z,n}(r, \varphi_0+\varphi)} \cos( \mu \varphi) \dd\varphi.
\end{align*}
Since the cosine is a symmetric function, it follows that in the presence of a reflection, $V_{n,\mu}(r) = \ee^{-2 \ii \mu \varphi_0} V_{n,-\mu}(r)$, i.e. the $+\mu$ and $-\mu$ components differ only by a phase factor.

The case in which the scattering coefficients for $+\mu$ and $-\mu$ components have the same absolute value may lead to the hypothesis that in such a case, no net OAM can be transferred as both scattering events happen with the same probability. However, this hypothesis clearly cannot be true as an arbitrary potential exhibiting only a mirror symmetry is not circularly symmetric and hence does not commute with the Hamiltonian. Therefore, Heisenberg's equation of motion together with Ehrenfest's theorem dictate that the net OAM has to change over time. The solution to this conundrum lies in interference effects.

\begin{figure}
	\centering\includegraphics{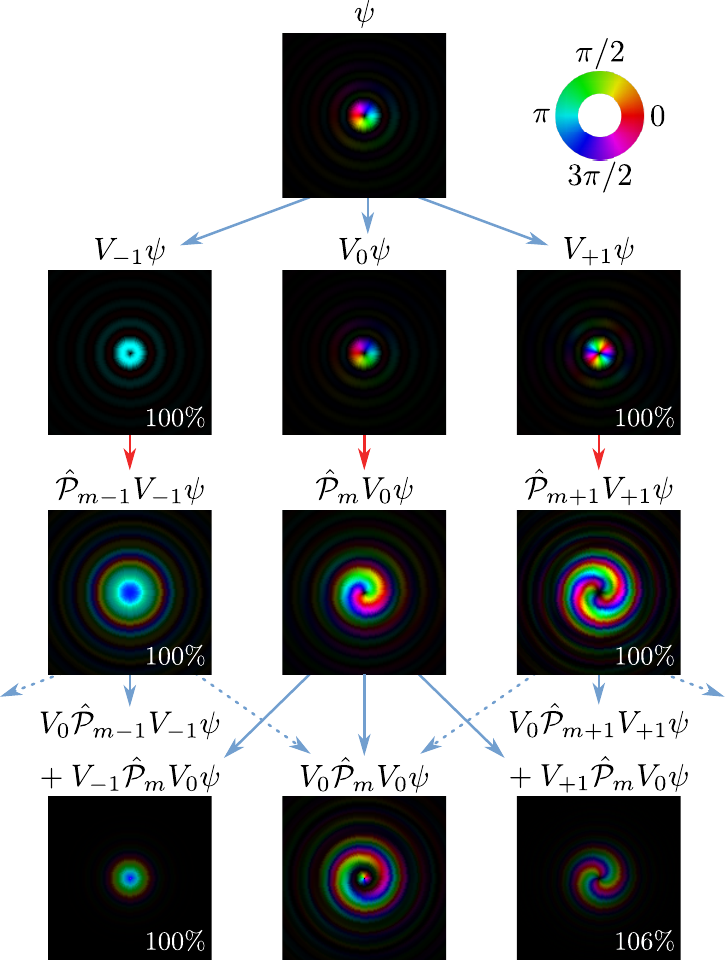}
	\caption{Schematic of the evolution of different OAM components upon transmission through two slices. The central column represents the (intially pure) vortex of order $m$ (the images show $m=1$), the left column shows the $m-1$ component, the right column shows the $m+1$ component. Blue arrows depict potential scattering while red arrows indicate Fresnel propagation (\op{\mathcal{P}}). Dashed arrows symbolize additional scattering contributions that are omitted for clarity. The insets in the left and right column depict the components' total intensities relative to $V_{-1}\psi$. The index for the slice number $n$ and the coordinates $r, \varphi$ were omitted. Brightness signifies amplitude, color signifies phase as depicted in the inset.}
	\label{fig:OAM_transfer_scheme}
\end{figure}

While the train of thought of equal probabilities is correct for single scattering, it breaks down when considering multiple scattering as depicted in fig.~\ref{fig:OAM_transfer_scheme} (for $\mu = \pm1$). There, it is clearly visible that after a single potential scattering event in the first slice, the $m-1$ and the $m+1$ components do have the same total intensity even though their phase structure is obviously different. The propagation behavior of the two components is also different, owing to the different orders of Bessel functions in eq.~\ref{eq:fn_after_prop}. However, as the Fresnel operator is unitary, the total intensity does not change during propagation.

The situation is different after the second slice, though. After the second potential scattering event, the $m+1$ component is given by the coherent superposition of two contributions. The first one stems from the portion of the beam that was first scattered with $\delta m=1$, then propagated as $m+1$, and then scattered again with $\delta m=0$. The second one stems from the portion that was first scattered with $\delta m=0$, then propagated as $m$, and then scattered with $\delta m=1$. The situation for the $m-1$ component is analogous, but not identical. Since propagation and potential scattering do not commute and the propagation is dependent on $m$, the interference patterns emerging from the coherent superpositions can be different for the $m-1$ and the $m+1$ components, thus leading to different total intensities as indicated in fig.~\ref{fig:OAM_transfer_scheme}. This, in turn, leads to a change of the OAM expectation value and, hence, to a net transfer of OAM, even though each individual potential scattering event is (quasi-)symmetric in amplitude for positive and negative $\delta m$.


\subsection{Radial Dependence}

Another interesting question is how the OAM transfer depends on the radius, which translates into the question of how the behavior of smaller and larger beams differs. For increasing $r$, larger and larger OAM transfers will become important. In fact, it is reasonable to assume that the dominant OAM transfer $\delta m\cdot \hbar\ne 0$ should scale proportionally to $r$. This can be deduced by comparing the mean atomic distance $a$ to the circumference of a circle with radius $r$. Since the mean distance is constant throughout the sample but the circumference scales linearly with $r$, the ratio of the two scales as $1/r$. For large $r$, this can be seen in a very crude approximation as the average period $p \sim a/(2\pi r)$ of a periodic oscillation of the potential as a function of $\varphi$. Thus, the frequency of this oscillation (which corresponds to the OAM transfer) is proportional to $1/p \propto r$. Consequently, one can expect that larger OAM transfers become more important with increasing beam size.

Obviously, for large $r$, there is also more room for variations, i.e. deviations from a perfect periodic oscillation with period $p$. Therefore, it can also be expected that the spread of possible OAM transfers should increase with increasing $r$. As an alternative argumentation leading to the same conclusion, one can invoke the uncertainty principle $\sigma[\varphi] \cdot \sigma[L_z] \sim \text{const.}$ \cite{NJoP_v6_i1_p103}: since angle and OAM are complementary variables, any localization in angle has to lead to a delocalization in OAM. Scattering on an atom produces a localized disturbance in the wavefunction with an initial angular extent of the order of $\sigma[\varphi]\sim a/r$. Thus, one can expect the standard deviation of the OAM to scale roughly proportional to $r$ as well.

\subsection{Expectation Value}

For some applications such as nanoparticle manipulation, the individual components of the OAM play only a secondary role compared to the expectation value of the OAM operator $\op{L}_z$, which corresponds to the total, net OAM of the beam. Directly in front of the $n$-th slice, this expectation value is given by
\[
	\expect{\op{L}_z}_{n-1} = \hbar \sum_m m \int_0^\infty |f_{n-1,m}(r)|^2 r \dd r,
\]
while behind the slice it is given by
\[
	\expect{\op{L}_z}_{n} = \hbar \sum_m m \int_0^\infty |f_{n,m}(r)|^2 r \dd r = \hbar \sum_m m \int_0^\infty |g_{n,m}(r)|^2 r \dd r 
\]
where the last equality holds due to Parseval's theorem (or, equivalently, due to the closure relationship of Bessel functions).

\section{Numerical Simulations}
\label{sec:sim}

In this work, we performed extensive numerical simulations for three amorphous model systems: \SiN, which is commonly used as support material and for phase masks; \Pt, which is commonly used as focused ion beam (FIB) protection layer and in absorption masks; and \FeB, a magnetic material used, e.g., in transformers, which could be interesting for EMCD. All simulations were carried out 5 times for randomly different atom arrangements to get an idea of the variations of the various results.

\begin{figure}
\centering\includegraphics{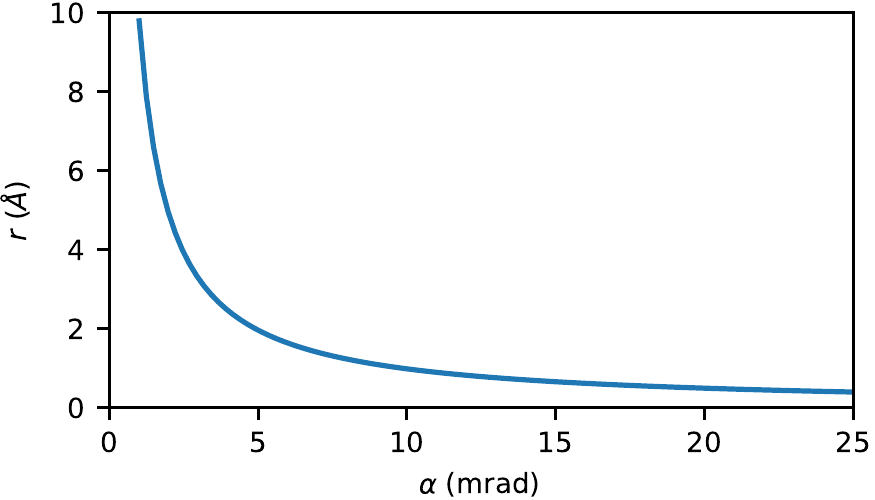}
\caption{Dependence of the $m=1$ beam waist radius $r$ on the convergence semi-angle $\alpha$ for \SI{200}{\kilo\electronvolt} electrons \cite{Loeffler2013}.}
\label{fig:beam_size}
\end{figure}

For all samples, a \SI{100x100}{\angstrom} area was simulated with \SI{512x512}{\pixel} using thicknesses in the range of \SIrange{0}{500}{\angstrom} with a slice thickness of \SI{2}{\angstrom}. All simulations were performed with a \SI{200}{\kilo\electronvolt} incident beam which initially was in an OAM eigenstate with $L_z = \hbar$. The convergence angles were in the range of \SIrange{1}{25}{\milli\radian}, corresponding to waist radii in the range of approximately \SIrange{10}{.4}{\angstrom} (see fig.~\ref{fig:beam_size}). For the sake of straight-forward interpretation, the experimental conditions were assumed to be ideal, i.e. the microscope lenses were assumed to be perfectly aberration-corrected and no broadening due to a partially incoherent source or motion of the atoms was included.

The atomic positions were generated at random, taking care that the overlap between adjacent atoms was as small as possible (i.e., rejecting atoms that were too close to already placed atoms). The used densities are summarized in tab.~\ref{tab:densities}.

All simulations were carried out using an in-house multislice code \cite{ACA_v68_i_p443,U_v131_i0_p39} based on the work by \citeasnoun{Kirkland1998}.

To evaluate the OAM components, the resulting wavefunctions $\psi_n(\rperp)$ after each slice were first transformed to a polar representation $\psi_n(r, \varphi)$ using a fixed $(r,\varphi)$ grid with \SI{256x1024}{\pixel}. Then, the transformation $\varphi \mapsto m$ was carried out by separately Fourier-transforming each line of constant $r$, yielding $\psi_n(r, m)$. Finally, the result was summed over the radius to obtain the total intensities
\[
	I_{m,n} = \int |\psi_n(r, m)|^2 r \dd r
\]
of each OAM component, which span the range from $m=-511$ to $m=512$. From these intensities, one can in turn calculate several physically relevant parameters such as the OAM expectation value
\[
	\expect{\op{L}_z}_n = \hbar \sum_m m I_{m,n}
\]
and the OAM variance
\[
	\sigma^2[\op{L}_z]_n = \expect{\op{L}_z^2}_n - \expect{\op{L}_z}_n^2 = \hbar^2 \left[ \sum_m m^2 I_{m,n} - \left( \sum_m m I_{m,n} \right)^2 \right],
\]
i.e., the squared standard deviation. Since all calculations were carried out for several randomly generated amorphous structures, we can also estimate the ``error bars'' associated with the physical quantities due to the fact that no two samples and no two position on a sample are identical.

\begin{figure}
\centering\includegraphics[width=\textwidth]{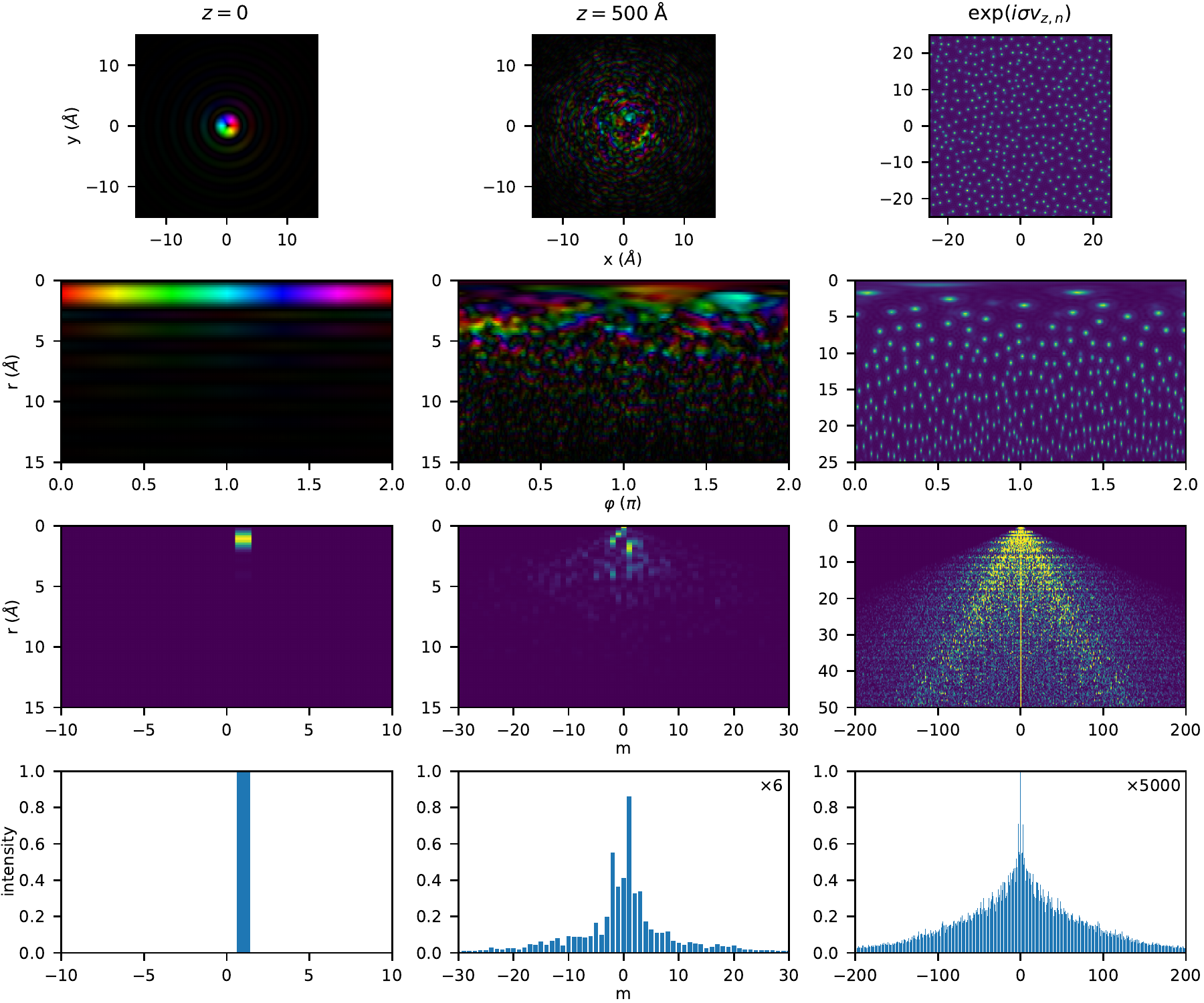}
\caption{Some examples of the data produced by the simulations and during the analysis. Left: wave-function at the incident plane of the sample, $z=0$; center: wave-function at the depth of $z=\SI{500}{\angstrom}$; right: scattering potential $\exp(\ii \sigma v_{z,n})$ of one slice. First row: data in Cartesian coordinates; second row: data in polar coordinates; third row: intensity of the $\exp(\ii m \varphi)$ components as function of $m$ and $r$; fourth row: total OAM intensities integrated over $r$. For the wave-functions shown in the first two rows, the phase is displayed as color (see fig.~\ref{fig:OAM_transfer_scheme}) and the amplitude is displayed as brightness. For the scattering potential in the first two rows of the right column, the argument of the complex exponential is shown. The total OAM intensities in the fourth row have been scaled as indicated. The incident beam was a pure $m=1$ vortex with a convergence semi-angle $\alpha=\SI{10}{\milli\radian}$ incident on the amorphous \ce{Pt} sample. In all cases, only a subset of the entire dataset is shown and the contrast was enhanced to improve visibility.}
\label{fig:example}
\end{figure}

Fig.~\ref{fig:example} shows some examples of the data produced by the simulations and during the analysis. In particular, it shows that, as predicted, the redistribution of intensity between different OAM components produced by the scattering potential is approximately symmetric but the resulting wavefunction has a distinctly non-symmetric OAM component distribution around the initial $m=1$ component. Under the given conditions, the $m=1$ component still exhibits the highest intensity, but the components $m \in \{-2, -1, 0, 2, 3\}$ have considerable intensities of the order of $I_{1,n}/2$. Therefore, their sum greatly exceeds $I_{1,n}$. Even higher orders in the range $-20 \lesssim m \lesssim 20$ also contribute non-negligibly, further emphasizing the broadness of the $m$ distribution. Interestingly --- though not surprisingly --- different $m$-components contribute strongly at different radii. In addition, the theoretically predicted linear increase of both the dominant $m \ne 0$ contributions as well as the $m$-spread in the scattering potential is clearly visible.

\section{Results}
\label{sec:results}

\begin{figure}
\centering\includegraphics[width=\textwidth]{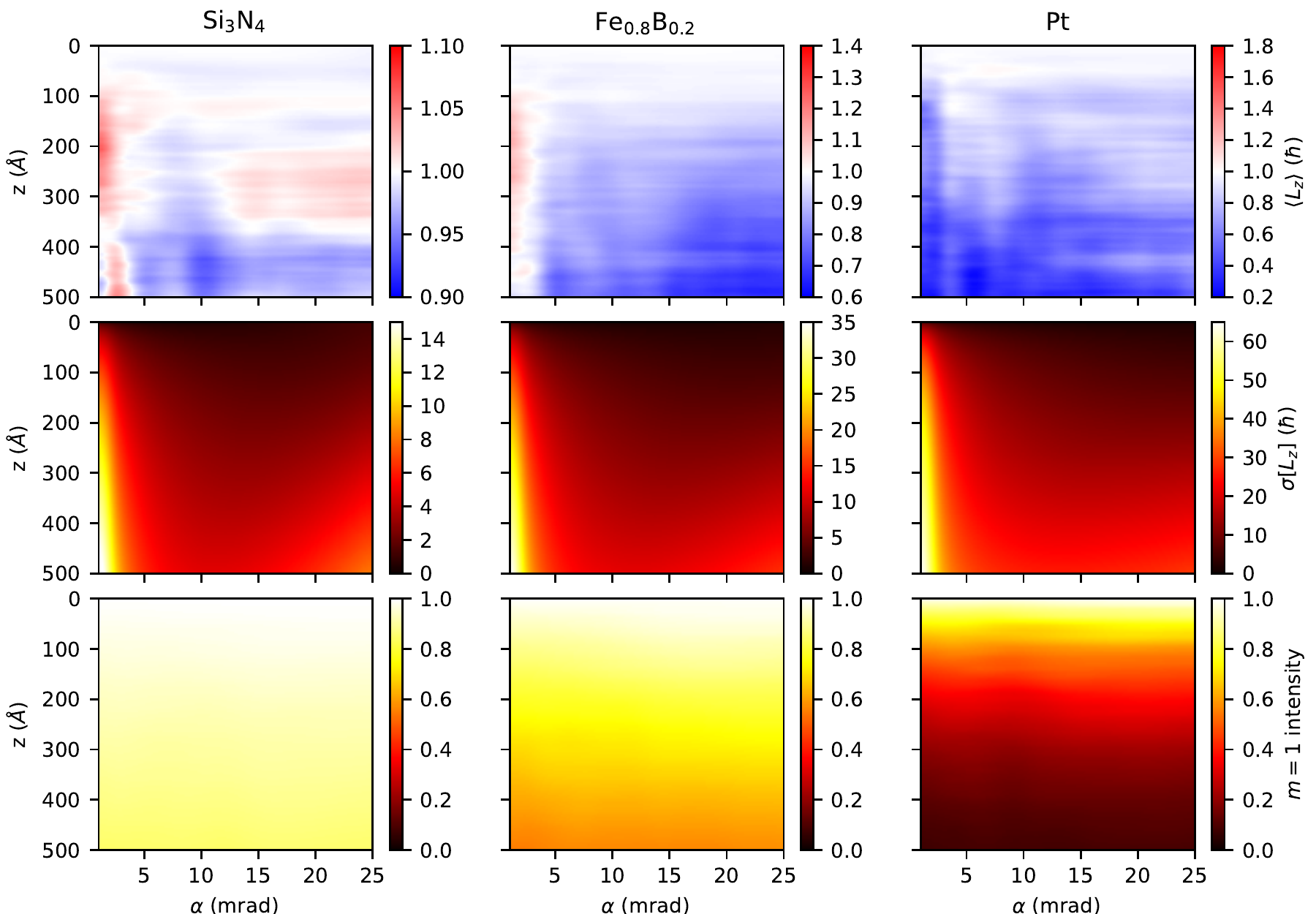}
\caption{Behavior of the OAM as a function of thickness $z$ and convergence semi-angle $\alpha$ for an incident $m=1$ vortex beam for three different samples. Left: \SiN, center: \FeB, right: \Pt. Top: OAM expectation value $\expect{\op{L}_z}$, middle: standard deviation of the OAM $\sigma[\op{L}_z]$, bottom: intensity of the $m=1$ component. Note the different color bar ranges. All data was averaged over 5 simulation runs.}
\label{fig:lzmean}
\end{figure}

Fig.~\ref{fig:lzmean} shows the dependence of several key quantities on the convergence semi-angle $\alpha$ (which is related to the beam size, see fig.~\ref{fig:beam_size}) of the incident beam as well as the thickness of the sample for the three simulated systems.

The most striking property is that while the overall features of the graphs are comparable between the three different systems, the numerical values differ greatly. Taking the maximum $\sigma[\op{L}_z]$ as an example, it changes from $\approx 15$ for \SiN over $\approx 35$ for \FeB to $\approx 65$ for \ce{Pt}. A similar trend is visible for $\expect{\op{L}_z}$. This phenomenon correlates nicely with the mass density of the three systems. Even though the atom density is comparable for \SiN and \FeB and is \emph{lower} for \Pt, the mass density \emph{increases} from \SiN over \FeB to \Pt (see tab.~\ref{tab:densities}), owing to the fact that \ce{Pt} atoms are much heavier than, e.g. \ce{Fe} atoms. Since heavier atoms generally scatter more strongly, it is logical that such systems produce stronger OAM deviations.

With respect to the changes of the expectation value $\expect{\op{L}_z}$, fig.~\ref{fig:lzmean} shows that the largest net OAM transfers occur for small convergence angles (i.e., large beams) and the smallest deviations occur in the range \SIrange{5}{10}{\milli\radian}, especially for small to medium thicknesses. This can be related to the size of the beam as it propagates through the sample. For small $\alpha$, already the incident beam is large compared to interatomic distances and it stays that way all throughout the sample. Thus, large $\delta m$ are possible from the very beginning of the propagation. For large $\alpha$, the diameter of the incident beam is small, but the beam size increases considerably during propagation. Thus, although initially only small $\delta m$ are viable, larger and larger $\delta m$ become dominant as the beam propagates further through the sample. Conversely, a beam with a mid-range $\alpha$ represents a good compromise between small initial size and small growth during propagation, thereby restricting the maximal significant $\delta m$ and, consequently, the variation of $\expect{\op{L}_z}$. A similar result was also found for classical EMCD (where vortex beams are generated during inelastic scattering and subsequently analyzed interferometrically) in crystalline samples \cite{M_v67_i_p60}.

For $\sigma[\op{L}_z]$ --- i.e. the OAM uncertainty or, equivalently, the spread of the $m$ distribution ---, the picture is very similar. Small $\alpha$ lead to a very large increase in $\sigma$ with thickness. For medium $\alpha$ in the range of \SIrange{7}{13}{\milli\radian}, $\sigma$ is smallest, while it increases again for large $\alpha$.

Interestingly, the $\alpha$-dependence is different for the $m=1$ intensity, which gives an indication ``how much'' of the original, incident beam structure actually is present at a given thickness. Fig.~\ref{fig:lzmean} shows that $I_{1,n}$ obviously decreases with thickness, but is mostly independent of $\alpha$. In other words: even though the net OAM and the $m$ distribution depend strongly on the beam size through the convergence angle and although there is complex multiple scattering going on back and forth between different $m$ components at different radii (as visible from fig.~\ref{fig:example}), the \emph{overall} intensity of the $m=1$ component seems to be fairly predictable.

\begin{figure}
\centering\includegraphics[width=\textwidth]{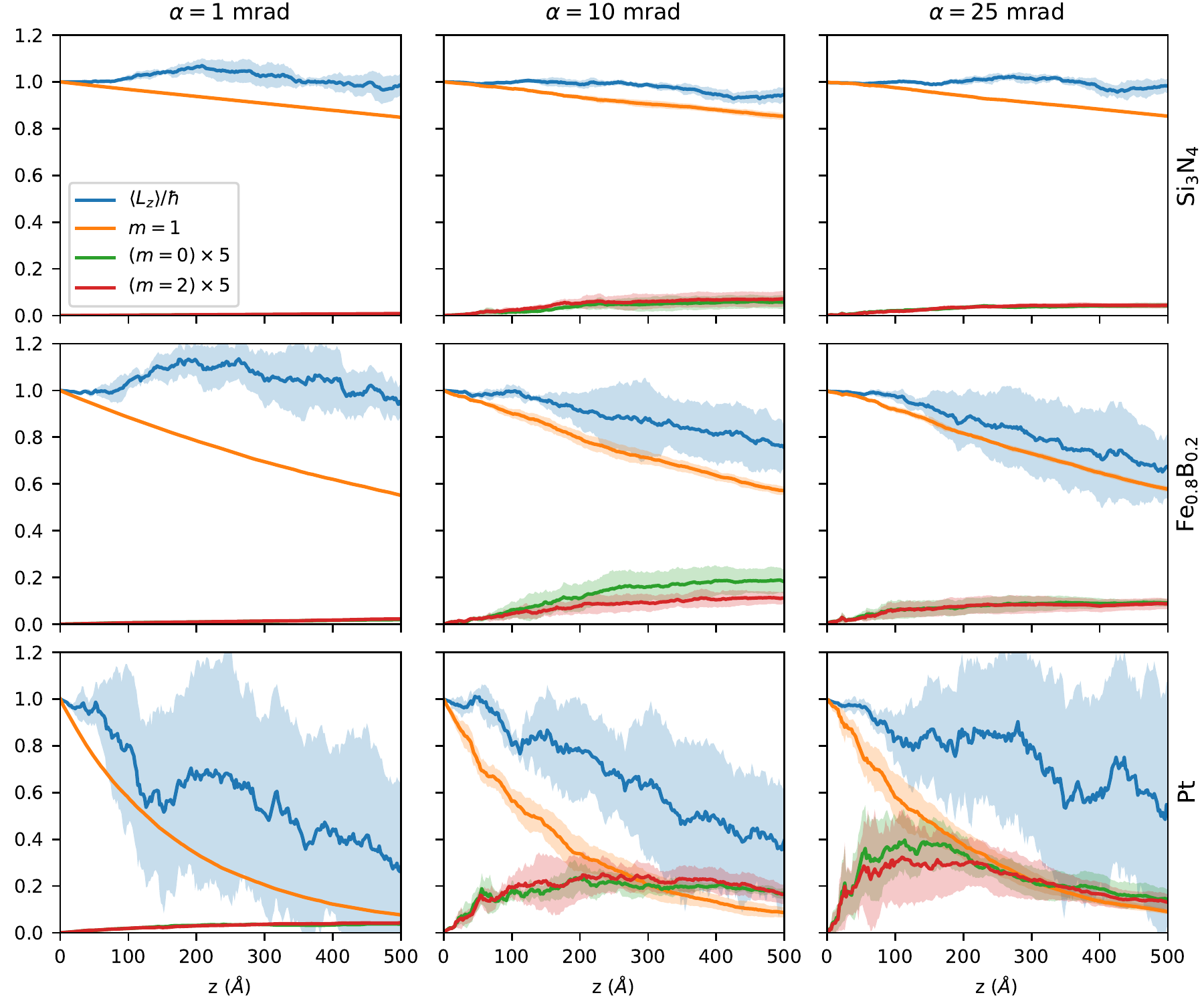}
\caption{Expectation value $\expect{\op{L}_z}$ and total intensities of the $m=0,1,2$ components for \SiN (top), \FeB (middle), and \Pt (bottom) for $\alpha=\SI{1}{\milli\radian}$ (left), $\alpha=\SI{10}{\milli\radian}$ (center), and $\alpha=\SI{25}{\milli\radian}$ (right) as a function of thickness $z$. The intensities of the $m=0$ and the $m=2$ components have been magnified by a factor of 5 as indicated in the legend to improve visibility. The shaded areas indicate one standard deviation as derived from 5 simulations.}
\label{fig:mrad}
\end{figure}

To investigate the intensity of different $m$ components as well as the expectation value in more detail, fig.~\ref{fig:mrad} shows graphs of the intensity of $m=1$ as well as the adjacent components $m=0,2$ and $\expect{\op{L}_z}$ for different convergence semi-angles $\alpha$ as a function of thickness. The adjacent components were selected because for applications that depend on the fact that the beam is in an $m=1$ eigenstate (such as, e.g. EMCD), typically close-lying other components are more difficult to separate than far-removed ones. As an example, an $m=100$ vortex would have practically zero intensity everywhere where an $m=1$ vortex is strong, thus making it easy to separate and block, e.g., by an aperture.

As before, the overall behavior of the curves is roughly similar for the $m$ components of all three studied systems, except for the scale of the thickness-dependence, which, again, is more dramatic for heavier specimens. Nevertheless, there are several noteworthy aspects visible in the graphs. In the first several \Ang, the decay of the $m=1$ intensity as well as the increase of the adjacent components is practically linear. This is to be expected as for a dominant $m=1$ component, the transitions $1 \mapsto 0$ and $1 \mapsto 2$ will be much more probable than the scattering $0 \mapsto 1$, $0 \mapsto -1$, etc. However, after several \Ang, all depicted components start to deviate from their linear behavior. The $m=1$ intensity decrease starts to slow down as soon as it reaches $\approx\SI{70}{\percent}$ of the initial intensity, while the $m =0,2$ intensities become almost constant somewhere in the range of \SIrange{10}{20}{\percent}. As the thickness increases, the $m=0,2$ components seem to asymptotically tend towards a similar behavior as the $m=1$ component, as is visible for \Pt at $\alpha=\SI{25}{\milli\radian}$ and --- to some degree --- already at $\alpha=\SI{10}{\milli\radian}$. Note that in all cases, the variation over several runs clearly indicates that the results are statistically significant, although the margin of error naturally is larger for larger mass density.

As already shown in fig.~\ref{fig:lzmean}, the decrease in $m=1$ intensity does not depend strongly on the convergence angle. However, the the increase of the adjacent components is influenced by the convergence angle. At the same time, the statistical uncertainty increases substantially for increasing convergence angles (i.e., smaller beam waists). This can be attributed to the fact that for sufficiently small beams (i.e., smaller than the inter-atomic distance), the propagation behavior is crucially dependent on the (random) relative position of the beam with respect to close-by atoms, whereas for large beams, the effect is averaged over many atoms.

Another interesting result can be found in the behavior and statistical variation of the expectation value $\expect{\op{L}_z}$. For \SiN, the deviation from $\hbar$ is marginal and fairly well contained in the statistical error. For heavier systems, the deviation from $\hbar$ become much stronger --- with a general trend towards decreasing $\expect{\op{L}_z}$ ---, but also the statistical variation between different simulations increases, up to the point that for \Pt at $\alpha=\SI{25}{\milli\radian}$, the deviation from $\hbar$ is no longer significant.

\section{Discussion}
\label{sec:discussion}

Whether the results presented here are encouraging or discouraging depends on the application at hand, the system under investigation, and the chosen experimental parameters.

If pure vortex beams are required, low mass-densities as in the case of \SiN and low thicknesses are definitely preferable in order to retain a high intensity in the $m$-component of the incident beam as well as little variation for different atom configurations. This also implies that holographic phase masks fabricated on a thick \SiN membrane can be subject to a considerable loss of mode purity.

If a high net OAM transfer is sought (e.g., in the case of nanoparticle manipulation), high mass-densities as in the case of \Pt as well as thick samples and medium convergence angles should be used. This ensures a large OAM transfer while retaining acceptable statistical variations for different atom positions.

It should be noted that real sample densities and scattering strengths may differ from the ones presented here, e.g. due to the use of different materials. In addition, the sample density is influenced by deposition and preparation parameters. However, the materials presented in this work span from fairly low to quite high mass densities and scattering strengths, thus giving a general insight into how arbitrary samples will behave.

All the simulations presented in this work were performed under ideal conditions, including no incoherent source size broadening (ISSB), no atom movement, and no lens aberrations. Both ISSB and atom movement would lead to an effectively different relative position between the beam and the atoms for each electron in the beam. This is conceptionally equivalent to the averaging over several random atom configuration as done in this study. Lens aberrations generally lead to a coherent broadening of the beam compared to the ideal case. While in such a situation the details of the amplitude and phase of the beam change, the overall results should be the same as those presented here when considering the appropriate beam size (see fig.~\ref{fig:beam_size}).

\section{Conclusion and Outlook}
\label{sec:conclusion}

In this work, we presented extensive simulations of the propagation of electron vortex beams through amorphous materials. To that end, we have rewritten the multislice approach into cylindrical coordinates to get some theoretical insight into the vortex propagation, such as the beam-size dependence of the redistribution of intensity between different $m$ components and the possibility of net OAM transfer despite the fact that the probabilities for transferring $\pm \delta m \cdot \hbar$ are (approximately) equal. In addition, we have also described the influence of the point group symmetry on the vortex propagation.

The numerical simulations were performed for the three amorphous model systems \SiN, \FeB, and \Pt for a wide range of convergence semi-angles and thicknesses. Besides corroborating the theoretical results, the numerical data allowed us to quantify the net OAM transfer, the spread of vortex components that is related to the uncertainty principle and, thus, the purity of a vortex state, as well as the intensity behavior of the most important vortex components. The results showed that in order to retain high purity upon propagation, low-mass-density samples with small thickness should be chosen, while large net OAM transfers can best be achieved in heavy, thick samples. In both cases, intermediate convergence semi-angles around $\alpha \sim \SI{10}{\milli\radian}$ proved beneficial.

The results presented in this work will allow theoreticians and experimentalists alike to choose the material for their studies with electron vortices more efficiently. Although this work does not completely replace full simulations for future studies, it does give some general insight into the propagation behavior of EVBs and makes predictions for a large range of systems and experimental parameters. As such, it promises to contribute to future enhancements not only of the fabrication but also of the applications of electron vortex beams.

\ack{T.S. acknowledges financial support by the Austrian Academy of Science (ÖAW) for the DOC scholarship and the ``Hochschuljubiläumsstiftung der Stadt Wien'' (project H-294689/2016).}

\referencelist[bibexport]

\begin{table}
\caption{Densities of the materials used in the simulations. The mass densities were used as reference. The atom densities were the ones used in the simulations.}
\begin{tabular}{lcc}
& Mass density (\si{\gram\per\centi\meter\cubed}) & Atom density (\SI[retain-unity-mantissa = false]{1e22}{\per\centi\meter\cubed}) \\
\SiN & 3.17 & 9.5 \\
\FeB & 7.18 & 9 \\
\Pt & 21.5 & 6.5 \\
\end{tabular}
\label{tab:densities}
\end{table}


\end{document}